# Mainstreaming of conspiracy theories and misinformation

N.F. Johnson[1,2,+,*], N. Velásquez[1,+], N. Johnson Restrepo[1,3], R. Leahy[1,3], R. Sear[4], N. Gabriel[2], H. Larson[5], Y. Lupu[6]
[1]Institute for Data, Democracy and Politics, George Washington University, Washington D.C. 20052
[2]Physics Department, George Washington University, Washington D.C. 20052
[3]ClustrX LLC, Washington D.C.
[4]Department of Computer Science, George Washington University, Washington D.C. 20052
[5]London School of Hygiene and Tropical Medicine, London WC1 7HT, U.K.
[6]Department of Political Science, George Washington University, Washington D.C. 20052
+ These authors contributed to this paper
* neiljohnson@gwu.edu

**Parents - particularly moms - increasingly consult social media for support when taking decisions about their young children, and likely also when advising other family members such as elderly relatives[1,2,3]. Minimizing malignant online influences[4,5,6,7,8] is therefore crucial to securing their assent for policies ranging from vaccinations, masks and social distancing against the pandemic, to household best practices against climate change, to acceptance of future 5G towers nearby. Here we show how a strengthening of bonds across online communities during the pandemic, has led to non-Covid-19 conspiracy theories (e.g. fluoride, chemtrails, 5G) attaining heightened access to mainstream parent communities. Alternative health communities act as the critical conduits between conspiracy theorists and parents, and make the narratives more palatable to the latter. We demonstrate experimentally that these inter-community bonds can perpetually generate new misinformation, irrespective of any changes in factual information. Our findings show explicitly why Facebook's current policies have failed to stop the mainstreaming of non-Covid-19 and Covid-19 conspiracy theories and misinformation[9,10,11,12,13], and why targeting the largest communities will not work. A simple yet exactly solvable and empirically grounded mathematical model, shows how modest tailoring of mainstream communities' couplings could prevent them from tipping against establishment guidance. Our conclusions should also apply to other social media platforms and topics.**

Facebook, the richest and largest social media platform, is struggling to prevent conspiracy theories and misinformation from moving into the mainstream on its own platform. Facebook is intervening more, and has a huge volume of material and users to moderate. But Fig. 1 suggests that the answer lies less in the amount of intervention to date, and more in where that intervention is -- and most importantly isn't -- targeted. Extended Data Fig. 1a shows the visible Facebook intervention, while Fig. 1 shows where these were targeted during 2020. Each node is a community (Facebook Page, see Extended Data Fig. 1) that has become enmeshed in the online health debate that in 2019 was focused around vaccinations[14] and now includes Covid-19's origin, severity and mitigation including vaccines. A directional link appears when community (node) A provides an explicit link at the Page level into community (node) B, hence opening up content feed from B and directing its users to look at B. Our data collection follows our published methodology[14] (see Methods and Supplementary Information (SI)).

Parents are turning increasingly to such in-built communities on Facebook to share concerns and seek advice from other community members on issues such as family health[1,2,3]. Mainstream parenting communities include those focused on children's wellbeing, education, discipline, breastfeeding, best milk choices, pregnancy, parental rights, homeschooling, home birthing, and special needs. Their average size is 57,895 users, but they range from 1,336,790 down to 25. Most focus on motherhood and most users appear to be women. While these Pages are publicly available, they often connect into private Groups and encrypted chats as shown in Extended Data Fig. 1b. Hence Fig. 1 is the visible skeleton behind which riskier follow-up discussions can take place, e.g. using non-approved treatments.

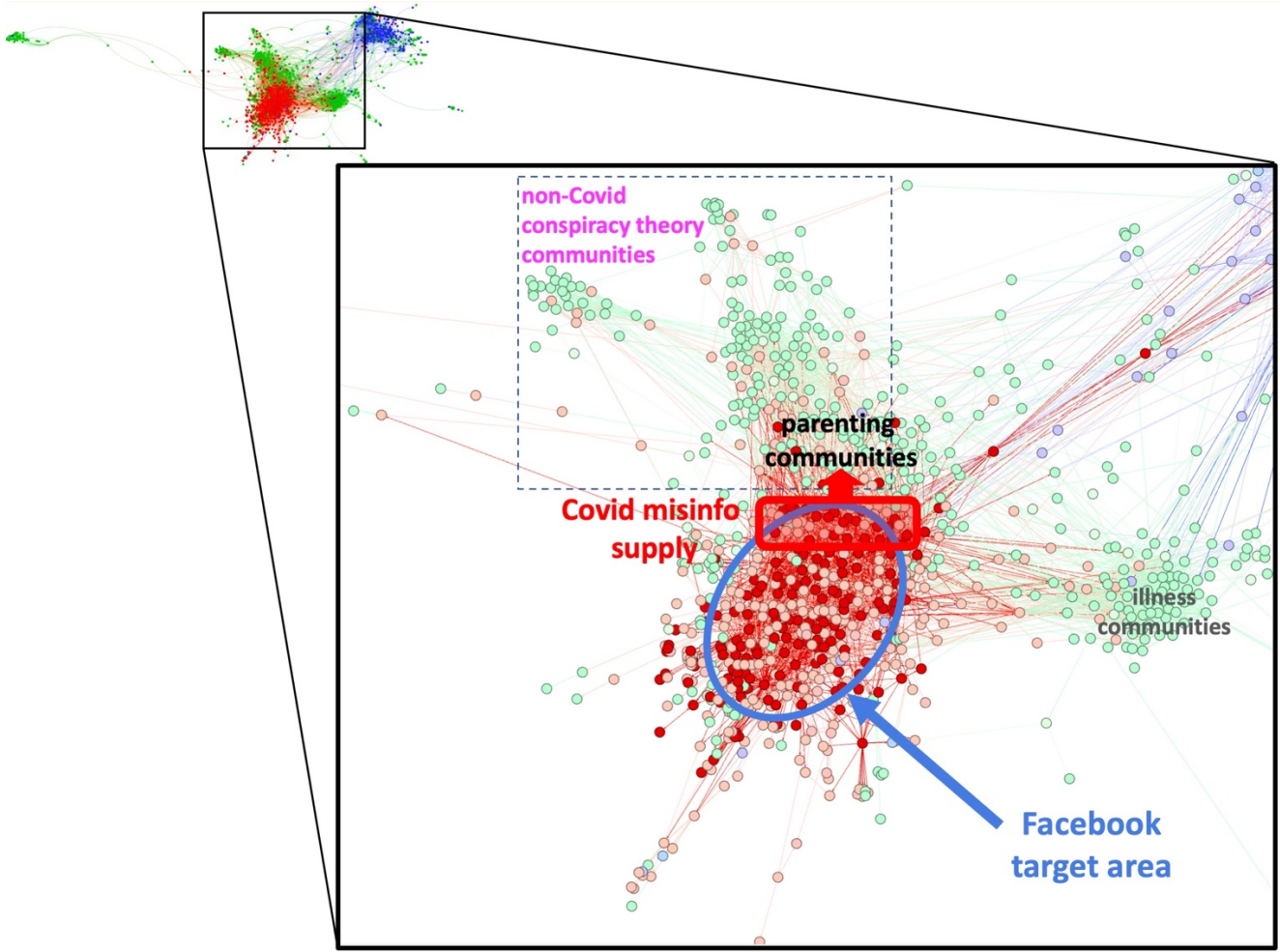

**Fig. 1: Facebook interventions (darker nodes) in the end-2020 Facebook ecosystem surrounding contentious health, containing >100 million users overall. Lighter nodes showed no Facebook intervention in summer 2020 when conspiracy theories and misinformation were taking hold. As in the pre-Covid 2019 version, each node is a community comprising 10-1,000,000+ like-minded supporters of a particular topic. Blue communities (nodes) support establishment health advice, Red communities (nodes) oppose it. Green communities (nodes) are not focused on such topics (see Extended Data Table 1) but have become linked to other communities that are. The ball-and-spring layout mechanism, ForceAtlas2, means that sets of communities (nodes) appearing closer together are more interconnected and hence likely have more shared content and users. Extended Data Fig. 2 and the SI explain how the network shapes in Figs. 1-3 can be explained quantitatively by the links that are present.**

Centola et al.[15] showed experimentally and theoretically that an online community can suddenly tip to an alternate stance in a reproducible way if there is a small committed minority of around 25%. This suggests that a connected community lying outside the main target ellipse in Fig. 1 could generate a cascade of tipping events across large portions of Fig. 1, e.g. dotted region. The untargeted dotted region alone includes ≈30 million users and contains most of the mainstream communities focused on parenting, as well as those promoting long-standing non-Covid conspiracy theories such as those related to chemtrails, fluoride and 5G. Figure 2 zooms into this dotted region, showing that during the Covid pandemic, there has been a visible shortening of the bond length between non-Covid conspiracy theory communities and parenting communities together with a reduction in the bond angle. However, this strengthening does not come from their direct bonding since there are very few direct links between the non-Covid conspiracy theory communities and the parenting communities. Nor, counterintuitively, is it mediated by communities against genetically modified foods (GMO) since these also have very few direct links to the parenting communities.

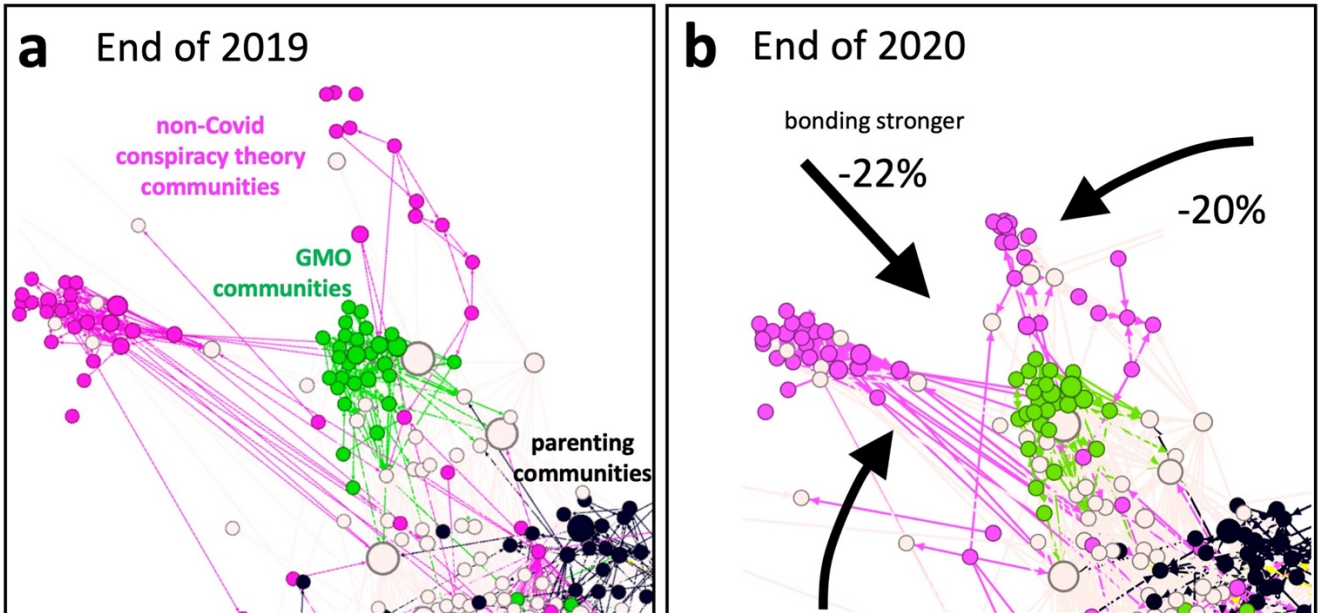

**Fig. 2: Parent--conspiracy-theory bonding strengthens during Covid.** Dotted portion from Fig. 1 before Covid (left) and during (right) using same scale. Distance between non-Covid conspiracy theory communities (e.g. 5G) and parenting communities shortens by 22% and angle reduces by 20%. But the key bond strengthening mechanism comes from alternative health communities (see Fig. 3a).

Instead, this increased bonding during 2020 between non-Covid conspiracy theory communities and mainstream parenting communities, comes via third-party communities focused on alternative health (Fig. 3a and Extended Data Figs. 3-5). The alternative health communities promote, discuss, and/or feature content about alternative cures and practices, as opposed to modern medical practice. This includes homeopathy, naturopathy, and spiritual healing. The Posts in these alternative health communities are not generally about conspiracy theories or vaccines, confirming that this is not these communities' overall focus or intention -- but deep in the content of the Replies to the Comments on the Posts, one can see new conspiracy theories and misinformation being generated that blend such themes from broader conspiracy theories, e.g. text in Fig. 3a combines narratives about World War III, 5G, vaccines, oil rigs, and vitamins. These alternative health communities tend to make the narratives more palatable, often inadvertently, by stressing the positive side of the untapped immune system and suggesting that parents have the power to provide their child with the best developed immune system possible. They show up to a 40% increase in their betweenness centrality (see SI) during 2020, which confirms their increased capacity during the pandemic to act as bonding conduits between non-Covid conspiracy theory and parenting communities. Such third-party bonding is common in complex biochemical molecules and is referred to as a 'superexchange' interaction.

Figure 3b reveals the remarkably robust inner engine that drives this misinformation ecology. Its main components sit mostly next to the parenting communities, in the red rectangle of Fig. 1, and are red (i.e. opposed to establishment guidance on vaccines and now Covid). This engine injects Covid-specific misinformation directly into the non-Covid conspiracy theory narratives from Fig. 3a. Its resistance to Facebook moderation stems from the facts that (1) these are not the largest red communities and hence are under-the-radar to moderators focused on size. But they do have the largest betweenness of all the communities in the entire ecology, and hence the greatest capacity to act as conduits (see Extended Data Table 2 and Fig. 6). (2) They are also highly interconnected among themselves, meaning that they each serve simultaneously as an effective individual conduit for the rest of the ecology in Fig. 1 and an effective conduit for each other. (3) They each tend to have a large imbalance between inward and outward links, meaning they act as net subscribers (yellow outer ring) or net broadcasters (gray outer ring). Thus they fit together like lock and key. (4) Several have administrators from across the globe, as shown, allowing them to easily shift their content across locations and inject local knowledge. (5) Several have a simultaneous presence on other platforms, where they direct their users to more extreme unmoderated discussions.

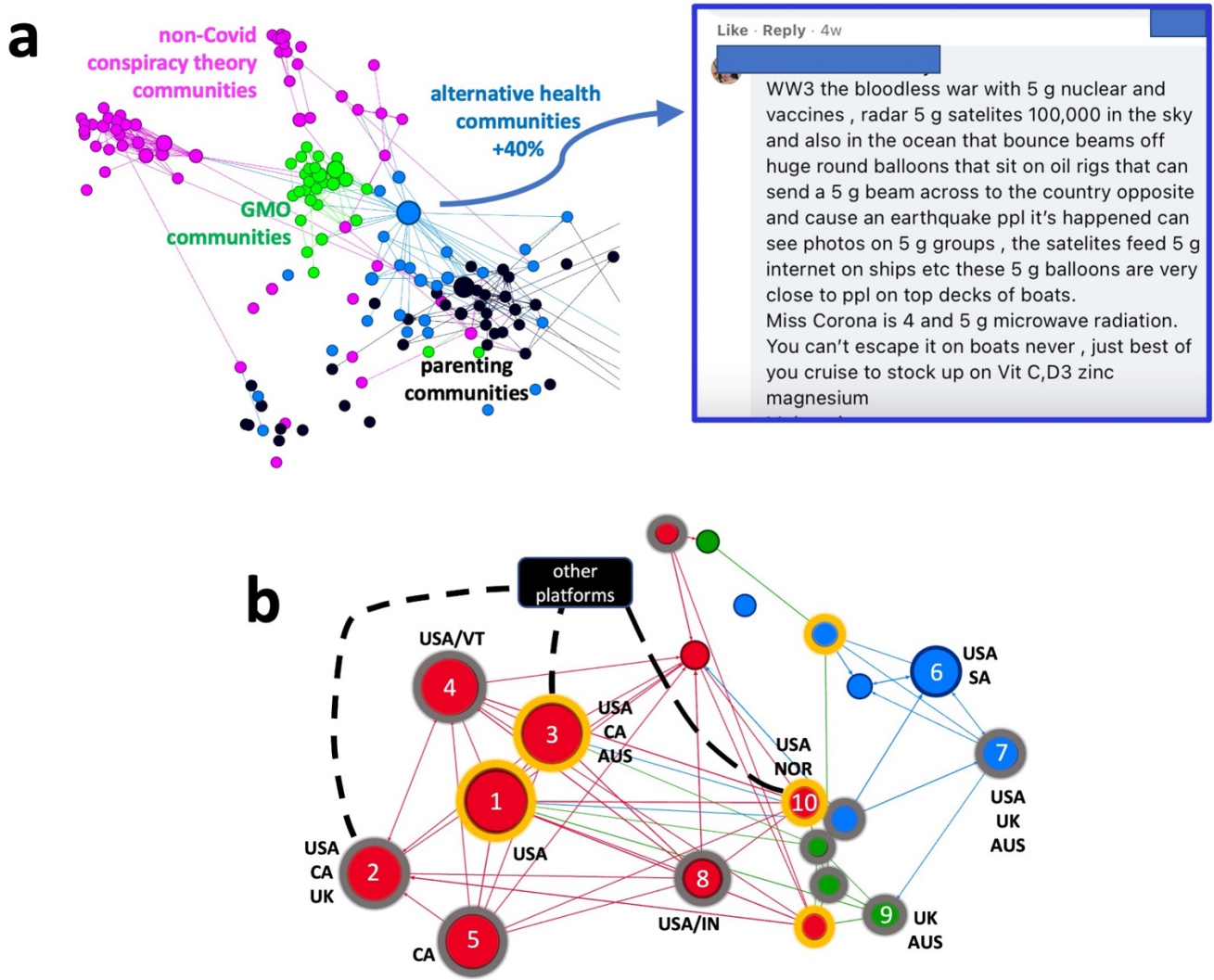

**Fig. 3: Key misinformation machinery. a:** Alternative health communities, which focus on positive messaging such as healthy immune system, provide the key bonding mechanism during 2020 between non-Covid conspiracy theory communities and mainstream parenting communities. Not because of their size, since these communities are not the largest (see Extended Data Fig. 6) nor because of any increase in links, but because of the huge increase in their betweenness centrality (shown as node size) and hence their ability to act as conduits, as a result of link rewiring during Covid. By contrast, neither GMO communities nor non-Covid conspiracy theory communities have many direct links to parenting communities. **b:** Top 20 communities by average betweenness centrality during Covid (see Extended Data Table 2 and SI), i.e. top 20 in ability to act as a conduit for (mis)information and conspiracy theories. Most of these red communities sit in the red box from Fig. 1, next to mainstream parenting communities. Yellow (gray) outer ring denotes overall emitter (receiver).

We have demonstrated experimentally the danger of such robust machinery if left untouched, by showing that existing conspiracy theories and misinformation can get repackaged as new in a way that can fool even subject matter experts (SMEs) familiar with the original pool of material. To mimic such repackaging, we employed a modified Turing test[16] using the open-source AI machine learning algorithm GPT-2 (see Methods). GPT-2 inputs existing narratives from across the communities that our SMEs had previously identified as containing health conspiracy theories and misinformation, and outputs seemingly new material without any top-down training or control -- just as might happen organically among the human users of the red communities in Fig. 3b. Our experiment shows that as long as the recycled texts are kept around the average length of human produced texts, the SMEs tend to see them as new human creations.

Our findings indicate that effectively combating online conspiracy theories and misinformation requires the identification of both their sources and conduits, and these may not be the largest or most prominent communities. The alternative-health-mediated bonding between conspiracy theory communities and mainstream parenting communities is outside Facebook's main intervention target area (Fig. 3a). The adjacent machinery (Fig. 3b) of the non-largest red (i.e., anti-vaccination) communities is highly resilient. Taken together, these phenomena explain why Facebook's current policies have failed to stop the mainstreaming of conspiracy theories and misinformation, and why targeting the largest communities will not work since they are not the major conduits of misinformation. Mainstream parenting communities are connected into conspiracy theories not primarily through vaccine safety, but by the control they feel they can have over the immune system of their children. Our findings complement the many valuable studies that focus on individual pieces of misinformation, individual communities, individual user accounts and news sources[4,5,6,7,8,9,10,11,12,13,17,18,19,20,21,22,23]. They also align with Ward et al.'s emphasis that effective messaging and intervention must account for the granular details, such as inequalities and differentiations between the actual communities to which people belong[24,25].

Our findings lead to a very simple, yet exactly solvable and empirically grounded mathematical model of this misinformation machinery (Fig. 4a) that suggests a new policy. It predicts that a modest change in the total coupling ($g_R + g_B + g_G$) of -82%, would move the current system across the tipping point (Fig. 4b, thick black arrow) such that parenting community concern will then decrease toward zero. Not only does this quantify the well-known but difficult policies of making red less concerning or blue more reassuring (i.e. make $g_R$ or $g_B$ more negative), it reveals a new one of using other greens as reassurance: i.e. make $g_G$ more negative. This amounts to reassuring parenting communities that other 'normal' communities are not concerned despite seeing the same online material, and could be easier to implement than trying to further restrict red or boost blue. Using the model to quantify the amounts, Facebook could achieve this by raising the profile of the Page links between parenting communities and other categories of green communities (e.g. pets) where the concern is currently low.

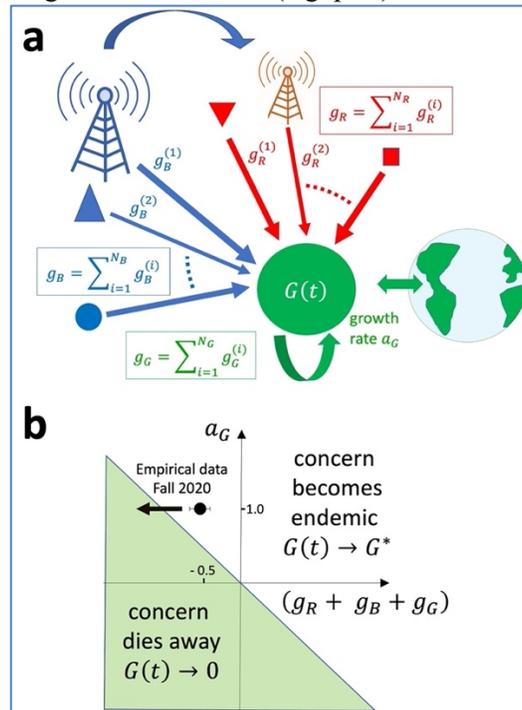

**Fig. 4: Misinformation tipping point. a: Blue communities and sources feed information to red and green communities. Red feeds its own interpretation to green. Green communities also feed each other (e.g. pet lovers to parents). $G(t)$ represents the concern shown by a specific green sector (e.g. parenting communities) as measured by volume of narratives. This simple model does not require specifying the nature of these information sources or the information itself, just the total strength of the net coupling ($g_R + g_B + g_G$). Assuming green concern goes down with more blue information and up with more red (mis)information, then $g_B < 0$ and $g_R > 0$. b: We extract the current numerical values by fitting the exact solutions to the actual activity in blue, red and green during 2020 (Extended Data Fig. 7).**

Limitations of our study include the fact that our approach to achieving understanding at scale necessarily washes over details of specific narratives. Also, online behavior does not automatically translate to offline behavior -- however it tends to be so for parents in these communities, since they regularly report back on prior advice they have implemented. Real-world impact of online conspiracy theories and misinformation can also be inferred from the civil engagement, activities and even protests parents report back on. In a non-health context, recent evidence of online-to-offline influence comes from the 2021 U.S. Department of Homeland Security national terrorism advisory[26]; recent online activity within a community of traders that produced huge movements in real-world stock and commodities[27]; and the use of Facebook to "foment division and incite offline violence" in Myanmar[28].

There are many very valuable existing approaches to managing online conspiracy theories and misinformation based on tracking, counteracting or correcting specific content, and inoculation[20]. However, scaling to the population level will still require knowledge of the system-level machinery in Figs. 1-3. Confronting specific narratives faces additional challenges: for example, the popular conspiracy theory that Covid vaccines feature hidden tracking devices, allowing personal information to be read from individuals' foreheads, derives from a false interpretation of a published article in a highly respected journal[29] that provides a proof-of-concept for this quantum dot technology. Efforts to label this as wrong science can hence exacerbate the level of concern (i.e. increasing $G(t)$ in Fig. 4). Also, our simulation in Extended Data Fig. 3c suggests that while inoculating against GMO narratives, for example, can result in many of the non-Covid conspiracy theory communities ending up farther from the mainstream parenting communities, a few may end up far closer. This brings to mind the phrase of behavioral psychologists[30] that the messenger (i.e. machinery of delivery) may matter more than the message.

**Methods**
Statistical analyses and details are given in the SI. While not perfect, our definition of nodes and links avoids the need for individual-level information, and makes the definition of each community unique. It also yields a visually manageable and interpretable network that is scalable to the population level, since each community contains of order 100,000+ users, and so the networks that we uncover of ≈1000 nodes (communities) account for 100+ million individuals. We focus on Pages of supporters of a given topic because they tend to foster a true sense of community among themselves. An AB link generally yields correlated activity (volume of narratives, see SI) between communities A and B. Though most communities are in English, users come from a wide range of countries. Since each community is self-policed, bot activity tends to be negligible. The exactly solvable model in Fig. 4, is of the form $\dot{G} = a_G(G_0 - G) + \sum_i g_{R_i}(R_i^* - G) + \sum_i g_{B_i}(B_i^* - G)$ where $G$ is the amount of concern in a particular parenting community, or a set, as measured by the volume of their narratives. $R_i^*$ and $B_i^*$ are fixed point values for red and blue. Taking $\epsilon_G$ as the deviation of $G$ from its possible fixed point $G^*$, the exact solution $\epsilon_G(t) = \epsilon_G(0)e^{-(a_G+g_R+g_B+g_G)t}$ using the definitions in Fig. 4a. Since $G$ has not reached its potential peak, then $\epsilon_G(0) < 0$ and so $a_G > g_R + g_B + g_G$ means $G$ increases over time toward $G^*$, while $a_G < g_R + g_B + g_G$ means $G$ decreases over time and becomes zero. Hence Fig. 4b follows. In the misinformation recycling experiment, the SMEs were told nothing to suggest that the narratives produced were not new. Yet they were all a machine generated blend of existing narratives using two versions of GPT-2 with two data preprocessing pipelines. All data contained only root posts and 10% of each dataset was held out for validation and perplexity calculation. One model ("antivax_gen1") was created with minimal preprocessing: anti-vax Facebook data collected through CrowdTangle, concatenating text attributes such as "title", "link text", "message", etc. The second model ("antivax_gen2") was created with a more targeted approach: anti-vax Facebook data from CrowdTangle, using only the "message" attribute. Additionally, we applied a regex before training to remove most URLs from the text. We used 15 different short text prompts (6 generic ones made of stopwords from NLP Python libraries, 9 incorporating topic keywords discovered by an LDA analysis used for previous research). For each prompt, we generated text with 3 different temperatures: 1, 1.1, and 1.2, so each GPT-2 model generated 45 text strings, for a total of 90 strings of generated text. Each string consisted of 500 tokens (words). For the survey, we bundled all 90 strings together into a dataset. From this dataset, we randomly selected strings such that no string's prompt appeared more than twice. We selected 15 strings with temperature 1, 3 with temperature 1.1, and 2 with temperature 1.2. 8 came from "antivax_gen1" and 12 came from "antivax_gen2". We truncated these strings to exactly 500 characters (approximately the average post length in authentic content) and placed them in an anonymous Google Form. None of the SMEs successfully identified all text strings as being repackaged by GPT-2, even though they all were. Four of the text strings proved very effective; two fooled 87.5% of participants, and two fooled 75%. Additionally, participants rating their confidence on the two most effective text strings (those which fooled 87.5% of participants) ranked their confidence, on average, as 4.13/5. The library used to bootstrap the GPT-2 models is available at https://github.com/huggingface/transformers#citation

**Acknowledgments:** We thank Rashmi Menon and Adwoa Brako for help with the diagram in Fig. 3b, and Om Jha for joint research on a triad variant of the model in Fig. 4. We are grateful for our use of CrowdTangle to access some of our data, through IDDP at George Washington University. N.V., R.L. N.J.R. and N.F.J. are funded by IDDP through the Knight Foundation. N.F.J. is also funded by AFOSR through grants FA9550-20-1-0382 and FA9550-20-1-0383.